\documentclass[]{interact}

\usepackage{hyperref}
\usepackage[T1]{fontenc}
\usepackage{multirow}
\usepackage{siunitx}
\usepackage{xcolor}
\definecolor{rev}{rgb}{0.0, 0.0, 0.8}

\usepackage{epstopdf}
\usepackage[caption=false]{subfig}

\usepackage[natbibapa,nodoi]{apacite}
\renewcommand\bibliographytypesize{\fontsize{10}{12}\selectfont}

\theoremstyle{plain}

\theoremstyle{definition}

\theoremstyle{remark}

\begin{document}

\articletype{RESEARCH ARTICLE}

\title{Using city-bike stopovers to reveal spatial patterns of urban attractiveness}

\author{
\name{Krystian Banet\textsuperscript{a}\thanks{CONTACT K. Banet. Email: kbanet@pk.edu.pl} Vitalii Naumov\textsuperscript{b} and Rafał Kucharski\textsuperscript{c}}
\affil{\textsuperscript{a}Department of Transportation Systems, Cracow University of Technology, Krakow, Poland, ORCiD: 0000-0002-9560-483X;
\\ \textsuperscript{b}Department of Transportation Systems, Cracow University of Technology, Krakow, Poland, ORCiD: 0000-0001-9981-4108; 
\\ \textsuperscript{c}Department of Transport \& Planning, Delft University of Technology, Delft, the Netherlands, ORCiD: 0000-0002-9767-8883}
}

\maketitle

\begin{abstract}
\color{rev}
We demonstrate how digital traces of city-bike trips may become useful to identify urban space attractiveness. We exploit their unique feature – stopovers: short, non traffic-related stops made by cyclists during their trips. As we demonstrate on the case-study of Kraków (Poland), when applied to a big dataset, meaningful patterns appear, with hotspots (places with long and frequent stopovers) identified at both the top tourist and leisure attractions as well as emerging new places.

We propose a generic method, applicable to any spatiotemporal city-bike traces, providing results meaningful to understand both the general urban space attractiveness and its dynamics. With the proposed filtering (to mitigate a selection bias) and empirical cross-validation (to rule-out false-positive classifications) results effectively reveal spatial patterns of urban attractiveness. Valuable for decision-makers and analysts to enhance understanding of urban space consumption patterns by tourists and residents.
\color{black}
\end{abstract}


\section{Introduction}
\color{rev}

Identifying urban places attractiveness and quantifying it is of high importance for policymakers, who can better design a city for city users; for the users, who may know which places are attractive; and for the local economy, which can find the optimal locations for their businesses. 
Yet urban space attractiveness is not at all easy to define, delimit and quantify. 
Cities are used by various groups, from daily commuters, through local visitors, business travellers, to tourists. 
Each with various activity patterns, needs and perceptions of urban space attractiveness. 
Collectively creating a complex spatial patterns, dynamically changing with emerging trends and fashions. 

In this study we demonstrate how a big spatiotemporal datasets of mobility traces may be used as a proxy revealing attractiveness of urban spaces. 
We contribute to the research stream where large sets of aggregated digital footprints are analysed to provide novel insights into how people experience the city \citep{girardin2009a}. By utilizing a relatively less exploited dataset (city bike traces) end exploring its unique feature (so-called stopovers) we reveal meaningful and valuable spatiotemporal patterns.

\subsection{Literature review}

In this section we first introduce the notions of urban attractiveness for tourists and local users along with methods to quantify and measure it. 
Then, we review a variety of recent methods leveraging on big datasets of digital footprints and their application to urban attractiveness. 
Finally we discuss city-bike systems, and a unique feature of digital-footprints left by city-bike users - stopovers.

\subsubsection{Urban space attractiveness}

Following the definition of \citep{biernacka2018classification}, the urban space is \textbf{attractive}, when one willingly wants to use it and spend her/his time there, and when this space corresponds with one’s individual needs, expectations and preferences.
Attractiveness of urban space is not at all easy to define, delimit and quantify \citep{boivin2019a}, with substantially different perception of urban space for tourists and locals \citep{kianicka2006locals}, notwithstanding both user groups are now better understood thanks to recent studies.
For instance, through the indicators to measure urban quality proposed by \cite{su10030575}, or with ‘City Love Index’, lately introduced by \cite{kourtit2020a}, which pinpoints attractiveness characteristics based on perceptions of urban quality by residents and their affinity with city life.
Residents' urban space consumption is associated mainly with their daily activities \citep{gonzalez2008understanding}, however its attractiveness for leisure purposes becomes increasingly significant \citep{THEES2020171}, better understood \citep{amanda2013} and quantified \citep{biernacka2020a}.

Likewise, the tourists' behaviour is better understood (for a thorough review we refer to \cite{cohen2014consumer}) through a studies where various segments \citep{stangl2020segmenting}, activity-based profiles \citep{fieger2019pull}, socio-demographic groups \citep{md2019a} or groups with specific needs \citep{lee2019a} are identified.
By means of tourist surveying \citep{jacobsen2019hotspot}, stated preference experiments \citep{gonz2019a}, or semi-structured interviews \citep{kianicka2006locals} attractiveness is typically related to a set of site-specific attributes \citep{estiri2020a} or individual visitors’ perceptions \citep{cracolici2009attractiveness, lee2019a}. Which, in turn, allows for a refined notion of tourism attractiveness at a national \citep{mitra2020a}, regional \citep{cracolici2009attractiveness}, city \citep{van2006attractiveness}, or site \citep{jacobsen2019hotspot} level. Which however, becomes challenging when within-urban attractiveness needs to be delimited \citep{zhu2020a}.

While attractiveness at macro-level can be identified via surveys, observing tourist movements plays a fundamental role in understanding their behaviour within the urban space  \citep{mckercher2008a}. To this end, tourists’ spatiotemporal behaviour – encompassing trajectories (movements between activities) \citep{zakrisson2012a} and stops (either at attractions, or to eat, rest, do shopping, etc.) \citep{caldeira2020a} - is analysed with implicit assumption that, in general, consumers of urban space spend more time in attractive spaces \citep{gehl2011a}.  

In such context, the movement along the multi-attraction itinerary can be observed \citep{huang2020tourists}, with participation time \citep{caldeira2020a} or time spent per bloc \citep{espelt2006a} used as attractiveness intensity indicators.
Early attempts to track tourists' movements using mental maps or self-completion diaries and surveys were usually time consuming and thus applied only on small sample sizes \citep{thimm2016a, keul1997a}.

\subsubsection{Digital footprints}

Digital footprints are now available in big volumes from numerous sources \citep{li_bigdata} which, coupled with a new kind of tourist that is avid for online content and predisposed to share information on social media, allows for a better understanding of tourist behaviour regarding their spatial distribution in urban destinations \citep{encalada2017a}.

Big volumes of data and its high availability seem to overweight limitations, mainly inherent selection bias \citep{SALASOLMEDO201813}. Consequently, big data in smart tourism \citep{li2017a}  contributes to understanding spatial patterns around urban tourist destinations and, for instance, to differentiate the overcrowded places from those with the potential to grow, allowing decision-makers to revisit planning and managing towards a sustainable ‘smart’ future.

User-generated social media content (photos on \textit{Twitter, Instagram, Flickr}, etc. or recommendations and reviews on \textit{TripAdvisor} or \textit{Booking}) have been widely explored in numerous studies \citep{hasnat2018a, miah2017a, oender2016a, li2018a, giglio2019a}. Recently \cite{marti2021taking} used Instagram data to reveal detailed picture of urban areas with most tourism-related activity – i.e. sightseeing, shopping, eating and nightlife - in Spanish cities.

While such data may reveal spatial patterns, it does not track the tourist movements, which requires geo-location data from personal devices (mobile phones) or vehicles (e.g. rental cars, scooters or bikes). GPS traces were used e.g. by \cite{girardin2009a} to provide insights into the attractiveness of urban space in NYC, by \cite{ORELLANA2012672} to explore visitor movement patterns in the Dwingelderveld National Park, by \cite{smallwood2012analysis} to understand distance decay in destination choice. \cite{zheng2017understanding}  used GPS to predict next destination within a Summer Palace in Beijing and \cite{ferrante2018general} tracked cruise passengers at the destinations.

\subsubsection{City-bike mobility traces}

Lately, bicycle sharing has become an increasingly popular around the world, making usage datasets big enough to study urban dynamics and aggregate human behaviour \citep{froehlich2009a}. City bike systems store records of trips with their origins, destination, and travel times in publicly available big databases, which allows for a rich understanding of mobility patterns \citep{cantelmo2019a}. Number of recent studies have used city-bike data e.g. to identify potential locations for new stations, estimate bicycle flows and usage, understand social and demographic context or predict usage in real-time \citep{caulfield2017a, eren2020a, frade2014a, imani2014a, salon2019a, tran2015a, wang_gender}. 
Paralller studies investigate how bike trips are affected by urban space factors such as: the number of retail stores and business offices near bike stations \citep{lin2020a}, demographic features \citep{wang_socio} or land-use \citep{kutela2019a}. 

How city-bikes are used by tourists was also studied. \citep{vogel2011a} identified that stations dominating between noon and afternoon were located directly near tourist hotspots,  \citep{brinkmann2020a} has shown differences in city bike usage between tourists and frequent users in Rio de Janeiro and Miami Beach. \cite{buning2020a} revealed different usage patterns between local residents and visitors, showing that visitors primarily use city-bike for leisure-based urban exploration, while residents’ use bikes mainly to commute. However, up to our knowledge, the unique feature of city-bike traces - stopovers was not exploited so far.

\subsection{Study overview}

Core of the proposed method lays in the concept of a \textbf{stopover}, a short stop made by a city-bike user during her/his trip.
Bicycle is not returned at the station, but stays with the user during a stopover.
Stopovers are typically short, since for longer stops users typically return bicycle to the docking station due to time-based fare scheme.
Stopover is not related to traffic, as we explicitly filter traffic-related stops e.g. at traffic lights. 
With such notion of stopover we may limit it to non-commuting trips, since commuters rent bikes to quickly reach the destination rather than to have stopovers. 

\begin{figure}
\centering
\resizebox*{10cm}{!}{\includegraphics{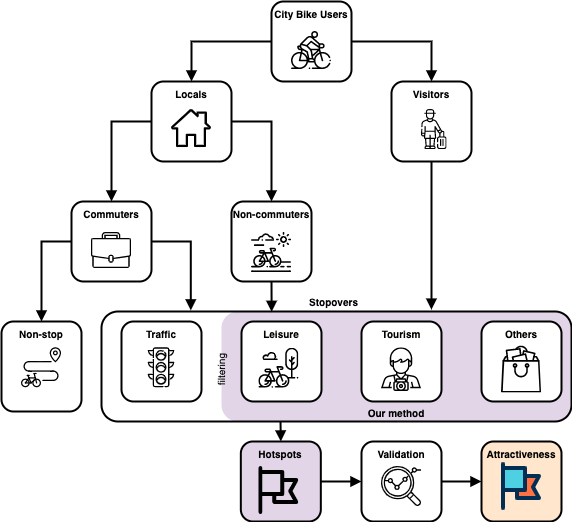}}\hspace{5pt}
\caption{City bike users classification. While local commuters stop mainly due to traffic, other user groups are likely to stopover during their trip. Both visitors and local non-commuters may stop at attractive leisure and touristic places, as well as to supply some their needs (e.g. shopping). Places with frequent and long stopovers (hotspots), after empirical validation, may be used as a proxy of urban space attractiveness.} \label{fig:Diag}
\end{figure}

Users may stopover for variety of reasons. The actual interpretation of stopover depends on the user type. We introduce \textbf{city-bike users} typology on figure \ref{fig:Diag}. Since the user details are missing we cannot distinguish commuters (using city-bike to reach their workplace), from local recreational users (having a weekend tour around green areas of the city), from business visitors (using city bike to reach the dinner with a client) and tourists (using city-bike to visit recommended tourist destinations). 

Nonetheless, we hypothesise that if city-bike user stops where she/he does not need to, it is mainly due to the place attractiveness, which may be either touristic, recreational, commercial or of any other kind. We further hypothesise that places where stopovers are frequent and long (following \cite{gehl2011a}), denoted urban \textbf{hotspots}, are attractive.

\begin{figure}
\centering
\subfloat[validating results of our method (rows) against actual space attractiveness (columns)]{%
\resizebox*{7cm}{!}{\includegraphics{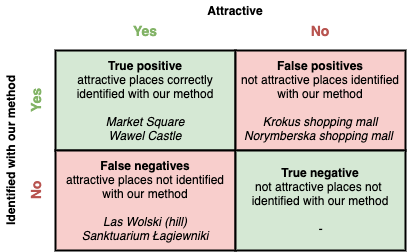}}}\hspace{5pt}
\subfloat[comparing our classification with classic static rankings]{%
\resizebox*{7cm}{!}{\includegraphics{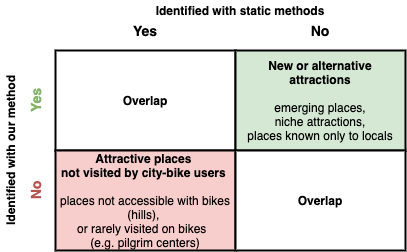}}}
\caption{Accuracy of our classification against actual attractiveness (left) and classic static methods (right).} \label{fig:matrix2}
\end{figure}

While we argue that identified hotspots are meaningful and valuable, we refrain from naively interpreting them as attractive urban hotspots. 
Like any other automated classification method, accuracy of our method is not perfect, as we illustrate with confusion matrix on fig.\ref{fig:matrix2}a.
The wrong classifications are either when our method failed to identify actually attractive places (e.g. places not accessible with bike, or outside of city-bike system) or when it identified places which are not attractive (where stopovers were not due to attractiveness, but for other reasons). 
We argue that \textbf{validation} of our results is straightforward, since each identified place may be examined against its true attractiveness relying either on the expert knowledge, other sources (social media or digital footprints from other sources) or a field visit.

\begin{figure}
\centering
\resizebox*{12cm}{!}{\includegraphics{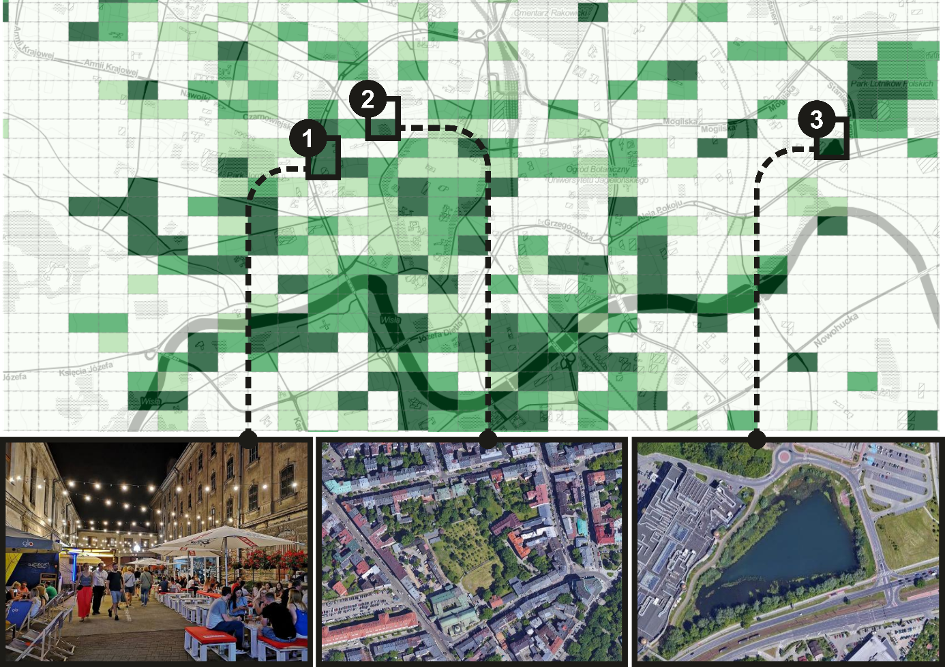}}\hspace{5pt}
\caption{Urban space attractiveness in Kraków, Poland. Identified hotspots were not only the classic points of any tourist itinerary, but also emerging places not mentioned in travel guides, e.g.: 1. the Dolnych Młynów pubs and clubs 2. Karmelicka str. gardens, 3. the Dąbski Pond.} \label{fig:f1}
\end{figure}

\subsection{Research problem, gap and contribution}

We contribute to the stream of research aiming to reveal spatial patterns of urban attractiveness. 
Identifying tourist hotspots and determining their dynamically changing attractiveness level is of crucial importance for decision-makers, who can now design the city better with attractiveness perception of residents and tourists in mind. 
Thus, in rapidly changing urban landscape, we need a dynamic method adapting to the recent trends and fashions of tourists \citep{dunne2011a} and residents \citep{kourtit2020a}.
Classic methods relying on expert knowledge and costly surveying fail to provide the detailed picture and are inherently static.
Recently, number of methods were proposed where big volumes of digital footprints were applied to reveal behaviour of residents and tourists in urban areas. 
Consequently, the delimitation of attractive urban spaces has become more detailed and underpinned with users' behaviour (observed via their digital footprints) and perception (understood thanks to surveying).
The objective of this paper is to demonstrate how this picture can be improved by using a new source of data and its unique features.

In this paper we exploit the potential of stopovers to reveal the spatial patterns. We hypothesise that the stopovers are related to the space attractiveness and verify it on a case-study. 
Nonetheless, observing stopovers is challenging. 
Stopovers cannot be read from social media data, even geotagged, which does not provide a participation time and, since it requires users action, content is not posted from all the places perceived as attractive. 
Detailed spatiotemporal digital traces are needed to reveal stopovers, and only the active travel modes (walking, scooters, bicycles, etc.) allow for unrestricted stopovers.
Cars are used by urban space consumers to a limited extend and most of attractive places are not accessible with car. Cars can be traced only to their parking spot and public transport passengers up to their bus stop.

Pedestrians exploring urban space are the least restricted to make spontaneous stopovers. 
However tracing pedestrians typically raises privacy issues and big volumes of personal mobile location data are not easily available. 
The privacy issues are partially overcome in a station-based systems, which does not contain sensitive personal data.
Consequently, the stopovers may be easily observed on a large scale only from the shared systems like scooters and city-bikes, which is their unique feature. In this study we exploit its potential.

Mining stopovers from detailed trajectories is not trivial. 
To this end we propose a novel method which allows first to identify stopovers from spatiotemporal trajectories and then to filter stopovers clearly not related to attractiveness. While results of the method needs to validated against external sources and local knowledge (as illustrated on fig.\ref{fig:matrix2}a) the revealed pattern accurately reproduced tourist hotspots of Kraków. While proposed method works with unlabelled data, the results may be refined when extra labels (sociodemographics or user-type) are available. 

Applying our method to the case of 35 thousands traces from Kraków, revealed a surprisingly meaningful and correct spatial pattern (fig.\ref{fig:f1}). Not only the main tourist attractions were properly identified, but also other insightful findings appeared. We identified a number of hidden gems, known only to locals, as well as newly emerging places, recently gaining popularity and often not yet listed on tourist websites.
Such places are unlikely to be timely identified via static studies, relying on expert knowledge (like in \cite{faracik2008a}), or surveys (like \cite{kianicka2006locals}) as we illustrate on fig.\ref{fig:matrix2}b, which is a central contribution of the paper. We demonstrate this with three examples on fig.\ref{fig:f1}. 

The study contributes to solving the problem of the tourist hot-spots identification by
using the exact statistical data against the traditional approaches that infer based on
unreliable information from the surveys and opinions of social media users. Identifies
a new feature of well-known digital footprint source, stopovers, widely available only
in city-bike traces.

The paper is organised as follows. In the next section we introduce a generic method to identify stopovers in mobility traces and apply it to the city bike datasets. We introduce a set of filters to calibrate the method before we synthesise the data on the spatial grid. In section 3 we illustrate the method using the example of Kraków, where stopovers identified in 35,000 bike trips yielded a grid that was validated against actual tourist hotspots. Finally, in section 4, we synthesise the results and discuss the potential applications and limitations of the proposed method.
\color{black}

\section{Method}

    We first formalize how stopovers are identified in the raw dataset, followed with a filtering rules, after which only meaningful stopovers remain in the dataset. Consequently, we aggregate the stopovers over a spatial grid and classify cells into four levels of attractiveness, the outcome of the method. The code to read the data from gpx-files, identify stopovers and compute attractiveness grid is publicly available
    \footnote{\url{https://github.com/naumovvs/city-bikes-analysis}}.

\subsection{Stopovers}

We analyse trip tracks, i.e., chronologically ordered sets of track-points:

\begin{equation}
    Track = \{TP_i\} \quad , \quad i = 1, \dots , N_{TP}
\end{equation}

, where each track point TP is defined as the time t and position:

\begin{equation}
    TP = \langle t, lon, lat \rangle
\end{equation}

For convenience, we use a dual definition, where trip becomes a set of trip segments:

\begin{equation}
    Track = \{ TS_j \} , \quad j = 1, \dots , N_{TS}
\end{equation}

and $TS_j$ is the $j$-th trip segment of the journey track defined by the couple of neighbouring track points $TP_o$ and $TP_d$:

\begin{equation}
    TS = \{ TP_o, TP_d \}
\end{equation}

From each trip segment we read: $t_s$ - the travel duration for the trip segment [h]; $d_s$ - the distance, calculated with haversine formula [km] and $v_s$ - the average travel speed, defined as the distance $d_s$ divided by the travel duration $t_s$ [km/h]. Consequently, the raw trips data now becomes:

\begin{equation}
    Trip = \langle ID, Track, t_{tr}, t_{idle}, d \rangle, 
\end{equation}
where $ID$ is the unique number identifying a trip in the dataset; Track is the reference to the object representing the GPS track as a set of track segments; $t_{tr}$ is the total travel time according to the track points data (the difference between the time moments when the last and the first track points in the track were read) [h]; $t_{idle}$ is the total idle time during the journey [h]; $d$ is the travel distance [km].

The total idle time is defined as the sum of travel durations for those travel segments for which the location has not been changed:

\begin{equation}
    t_{idle} = \sum _{TS_{idle}} ts , TS_{idle} = \{ TS_j:d_{j}=0 \} , \quad j = 1, \dots , N_{TS}
\end{equation}

, where $TS_{idle}$ is a set of all segments within the trip that have zero travel distance.

Stopovers, central element of the proposed method, are a set of consequent travel segments with null distance:

\begin{equation}
    Stopover = TS_{idle} \{TS_k , TS_{k+1} ts \}  , \quad 1 \leq k \leq N_{TS}. \label{eq7}
\end{equation}

\begin{figure}
\centering
\resizebox*{8cm}{!}{\includegraphics{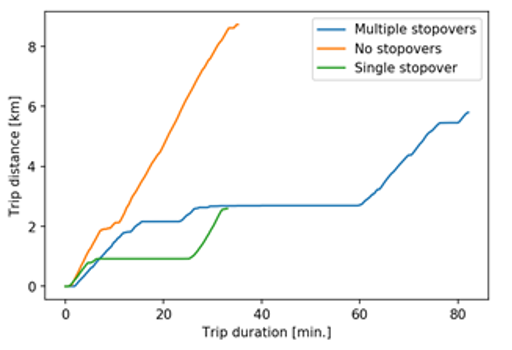}}
\caption{Typical spatiotemporal traces of city bike trips with various number of stopovers. Commute trips typically are with no stopovers (orange) while leisure or touristic trips are more relaxed and are often intermitted with stopovers.} \label{fig:f2}
\end{figure}

Note that, according to the definition, more than one stopover could be defined within a trip, if travel segments with zero distance are not consequent segments. We illustrate three selected rides with various number of stopovers on fig.\ref{fig:f2} .

\subsection{Filtering}

To make sure that identified stopovers are meaningful, we applied the following pre-processing sequence:

\begin{enumerate}
    \item 	The dataset was first cleansed of corrupt records related to signal failures in GPS transmitters. Trip data were eliminated if the GPS outage lasted at least five minutes. 
\item	After the first filtering stage, the sample continued to contain many trips with an average speed of 0 km/h. Therefore, we decided to eliminate all trips with a null duration or distance, which were most likely related to situations where a bike was unlocked, but not taken out of the stand, and then locked again, e.g. because of a technical problem. 
\item	Subsequently, we removed very short trips, where a technical problem was discovered soon after the bike was rented and the user decided not to continue with the trip. Based on \citep{naumov2020a}, a 50 m threshold was used.
\end{enumerate}

On such filtered sample we identified stopovers using the method proposed above. For each recorded trip, stopovers were identified with (eq. \ref{eq7}) with their location and duration. Yet, a heatmap visualization of obtained stopovers revealed a need for further filtering. 
To this end we applied a second stage of filtering using spatial metadata from OpenStreetMap as follows:
\begin{enumerate}
\item	Bike rental/return stops.\\
Most evident was the need to filter stopovers in the proximity of BSS stations, where what our method identified as stopovers were in fact unlocking and locking the bike and checking its technical condition. We identified threshold of 7m around BSS station to efficiently eliminate trip starts and ends from stopovers. We explored the cumulative number of stops as a function of distance from origins and found a natural cut-off point at 7 meters, after which the number of stops stabilized \citep{naumov2020a}.
\item	Stops at traffic lights.\\
Obviously, most of times when city bike users stop is not for sight-seeing, but at traffic. This had to be filtered with care to obtain meaningful results. Importantly, in the vicinity of most of Kraków’s tourist attractions there are no traffic lights, so we could safely assume that stopovers around traffic lights are not due to attractiveness. Our analysis revealed that the number of stops stabilized at a radius of 30 m; adopting a greater radius would lead to discarding stopovers unrelated to the presence of a pedestrian crossing or intersection.
\item	Railway crossings stops. \\
Despite there is just handful of single-level railway crossings in Kraków, stopovers in their vicinity (definitely non-attractive places) biased the emerging picture. Since those were just few points, it was easy to identify them and manually filter at 30m threshold.
\item	Short, traffic related stopovers.\\
While above filters were spatial, we decided to apply also a temporal filter, which we found efficient in filtering short, traffic related stops. Namely, we found that vast majority of stopovers below 30s were around unsignalized pedestrian crossings. So we filtered stopovers in vicinity of pedestrian crossings shorter than 30s 
\end{enumerate}

After above stages of filtering, the meaningful spatial patterns started to emerge, with heatmaps now clearly resembling tourist attractiveness, rather than a traffic map. Notably, in the above we did not need to map-match traces, which makes the methods light and generally applicable. 

\subsection{Aggregation}

For meaningful and quantifiable visualization, we divided the analysed area into a number of rectangular fields of a given size $S$. We used a spatial grid that can be represented as the following matrix:
\begin{equation}
    	Grid= \vert \vert Field_{ij} \vert \vert ^{S \times S},i=1\dots S,j=1\dots S,
\end{equation}
where $Field_{ij}$ is the rectangular field representing the part of the analysed area; $S$ is the grid size (the greater number of cells, the more detailed results).
For each field we get:

\begin{equation}
    	Field=\{n_{st},t_{st},t_{F} \} ,	
\end{equation}

where $n_s$ is the total number of stopovers in the field; $t_{st}$ is the total duration of all the stopovers in the field \textit{[h]}, and $t_F$ is the mean stopover time (calculated only for cells with more than three records):

\begin{equation}
t_F =  \begin{cases}
        0 & n_{st} <3\\
        t_{st} / n_{st} &\text{otherwise}
    \end{cases}
\end{equation}

All being potentially useful to reveal the spatial attractiveness. As we demonstrated for our case-study in the next section, we decided to base attractiveness on the mean duration of stopovers rather than their number or total duration. To sharpen the emerging picture, we classified grid cells into four attractiveness classes, representing quartiles of mean stopover duration ($t_F$), where 4-th class represents highest attractiveness and 0-th lowest. The final outcome of the method is a spatial grid with rank (from 0 to 3) of urban space attractiveness for each field.

\section{Results}

We illustrate the proposed method with the case of Kraków, Poland,  one of emerging tourism centres in Europe. With its rich history and unique cultural heritage, the city has attracted a growing number of tourists in the last decade. For Kraków, tourism is not just an important source of revenue, but also a major social phenomenon that shapes its urban identity. Tourism in Kraków has long been concentrated around the historic city centre (Old Town, Wawel Castle, Kazimierz), i.e. the urban complex inscribed in the UNESCO World Heritage List. Yet now it extends to neighbouring areas, such as Kleparz, Krowodrza, Zabłocie, Stare Podgórze, and Nowa Huta, due activities aimed at decentralizing tourist traffic \citep{tracz2018a}. Visitors are becoming more and more heterogenic, spanning from John Paul II related pilgrims, to city-breakers focused on nightlife, from high-school pupils visiting their national royal Castle for the first time, to frequent visitors looking beyond the top-10 sights. These dynamics and diversity yield rapid and complex patterns which are hard to trace and quantify.

We used the data from the local city bike system, ‘Wavelo’, established in 2008 and gradually expanded until 2019 when its area covered majority of the city. The records covered one week of the high tourist season of 2017, i.e., from 31st of May to 7th of June, when weather was auspicious for bike traffic and recreation, with mean temperatures between 16 and 21$^{\circ}$C and barely any rainfalls. The dataset composed of 34,969 tracks of Wavelo city bike users was sufficient to obtain clear and meaningful patterns. The high-fidelity and pre-processed mobility traces in the gpx-format were obtained from GPS transmitters attached to every Wavelo bike .

The first step was to cleanse the dataset provided by the city bike system. First, we eliminated all data corrupted by GPS transmission failures (a total of 5,946 trips). At the second stage, 40 trips with a null duration and 635 trips with a null distance were removed. The third stage, which involved filtering out short trips, identified 421 trips with a distance shorter than 50 m. Once these were eliminated, the final sample consisted of 27,927 routes.

The number of stopovers in the cleansed sample was 54,143, with a mean stopover time of ca 80s. After the last filtering stage, the number of stopovers dropped to 5,791, while their mean duration rose to a little over 6 minutes; only 25\%, however, were longer than 5 minutes 35 seconds (tab. 1). The largest drop in the number of stopovers was recorded after the first step, which involved eliminating those within a radius of 7 m from the trip’s origin. Further decreases were less steep, but a clear relationship could still be observed between the increase of the mean duration and the lower number of stopovers in the sample.

\begin{table}[]
\centering
\resizebox{\textwidth}{!}{
\begin{tabular}{lSSSSSSSSS}
\toprule
     stage                  &               \# stops              & &    & {time [s]}   &     &  &  &  &        {sum} \\ 
                       &                                  & {mean}   & {std}    & {min} & {25\%} & {50\%} & {75\%} & {max}   & [h] \\ \midrule
raw      & 54143                            & 79.17  & 291.63 & 1   & 15   & 25   & 65   & 27350 & 1190.77     \\
1st                    & 9639                             & 280.93 & 559.30 & 1   & 85   & 90   & 255  & 8290  & 752.20       \\
2nd                    & 6277                             & 335.83 & 647.13 & 1   & 85   & 155  & 275  & 8290  & 585.56      \\
3rd                    & 6219                             & 337.74 & 649.78 & 1   & 85   & 160  & 275  & 8290  & 583.45      \\
4th                    & 5791                             & 361.80 & 667.09 & 31  & 85   & 170  & 335  & 8290  & 582.00    \\ \midrule
\end{tabular}}
\caption{Number of stopovers and their statistics in the subsequent filtering stages.} 
\end{table}

After each filtering stage, we visualised the dataset for verification of the obtained results. The unfiltered sample was dominated by punctual stops near rental stations. The first filtering stage allowed us to identify stops that were not related to the trip origin or destination. After the first stage, the map still contained punctual hotspots related to stopovers at junctions and level crossings which were successfully filtered with subsequent filtering stages. Finally, most stopovers were identified in the touristic and leisure places like Vistula Boulevards, Old Town and the Main Market Square. 

\begin{figure}
\centering
\resizebox*{12cm}{!}{\includegraphics{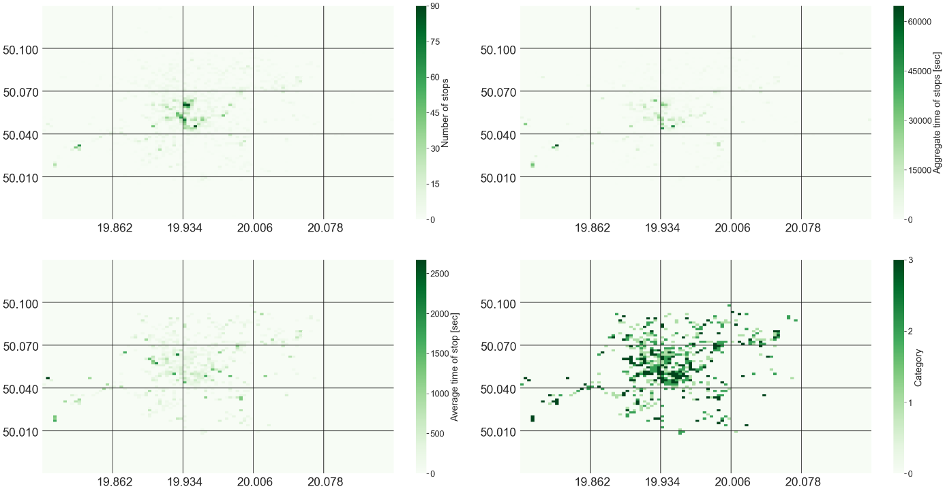}}
\caption{Values on the 100x100 grid of a) number of stopovers, b) total stopover time, c) mean stopover time and d) mean stopover time in four classes. Clearly the last one showing most evident patterns resembling the actual structure of Kraków’s tourist and recreational attractiveness.} \label{fig:f4}
\end{figure}

In order to quantify results, we created a grid where each field was assigned the corresponding number of stopovers and total stopover time (eq. 9). In the case at hand, the area delimited by the outer geographical coordinates of the recorded stopovers was divided into 10,000 fields of the same geographical longitude and latitude, i.e., 0.0015$^{\circ}$ × 0.0036$^{\circ}$. Figure \ref{fig:f4} shows values on the grid of: number of stops (a), total stopover time (b), mean stopover time (c) and mean stopover time in four classes (d). Based on the emerging patters, it was evident that the mean stopover time provides a meaningful proxy to identify urban hotspots. Other ones yielded both false-positive as well as true-negative errors, where identified hotspots were not attractive and attractive hotspots were not identified, respectively.

\begin{figure}
\centering
\resizebox*{8cm}{!}{\includegraphics{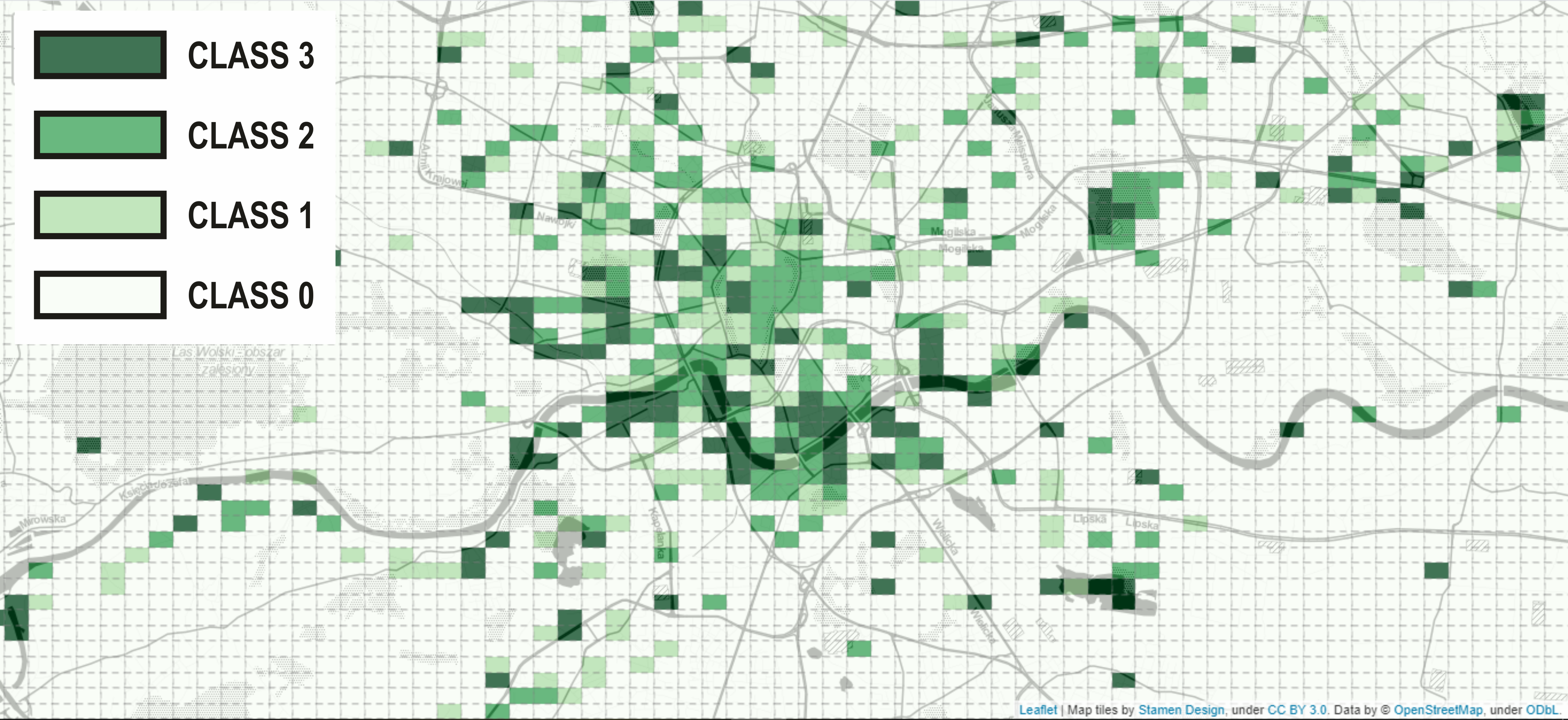}}
\caption{Four classes of urban attractiveness in Kraków based on the mean stopover times of city bike users.} \label{fig:f5}
\end{figure}

Most fields in the attractiveness identification grid, i.e., 94.74\%, have a rating of 0, but the most attractive areas of the city, such as the Old Town, the Vistula Boulevards, the Benedictine Abbey in Tyniec, the Kolna kayaking centre, the Polish Aviators’ Park, Bagry Lake or Nowa Huta Lake, achieved a high attractiveness score. Table 2 shows the number of fields in each attractiveness category and their percentage share in the total number of fields in the attractiveness identification grid.
 
\begin{table}[]
\centering
\begin{tabular}{lSS}
\toprule
class & {\# fields} & {share}  \\ \midrule
0 (lowest)           & 9474             & 94.74\%                             \\
1                    & 183              & 1.83\%                              \\
2                    & 183              & 1.83\%                              \\
3 (highest)          & 160              & 1.60\%      \\ \midrule                       
\end{tabular}
\caption{Attractiveness classes of attractiveness identification grid fields.} 
\end{table}

\subsection{Validation}

\begin{figure}
\centering
\resizebox*{8cm}{!}{\includegraphics{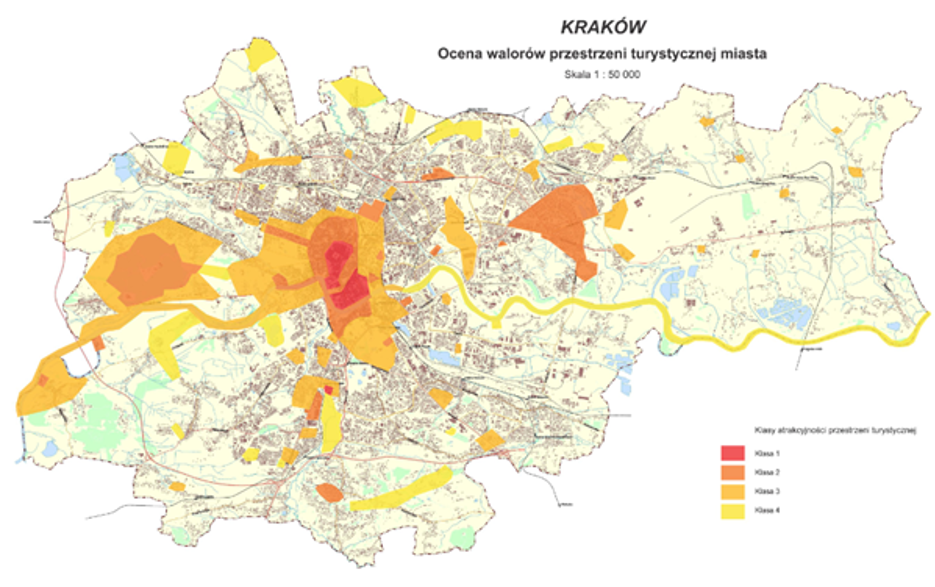}}
\caption{Four classes of tourist attractiveness in the official policy of Mayor of Kraków \cite{faracik2008a}.} \label{fig:f6}
\end{figure}

Since the proposed method is explorative and aims for more complete identification of previously unrevealed attractive urban hotspots, its’ validation is not straightforward. Nonetheless, to demonstrate its’ capabilities we validated our results against typical sources of tourist attractiveness. \cite{faracik2008a} evaluated urban space in terms of its tourism attractiveness, which was later adopted as an official and the latest tourism policy by the Mayor of Kraków. In the comprehensive study, they relied on their expert knowledge to select the natural, cultural, accommodation and services factors in the assessment process and, similarly to our study, classified city space into four classes, as shown in fig. \ref{fig:f6}. 

\begin{figure}
\centering

\resizebox*{10cm}{!}{\includegraphics{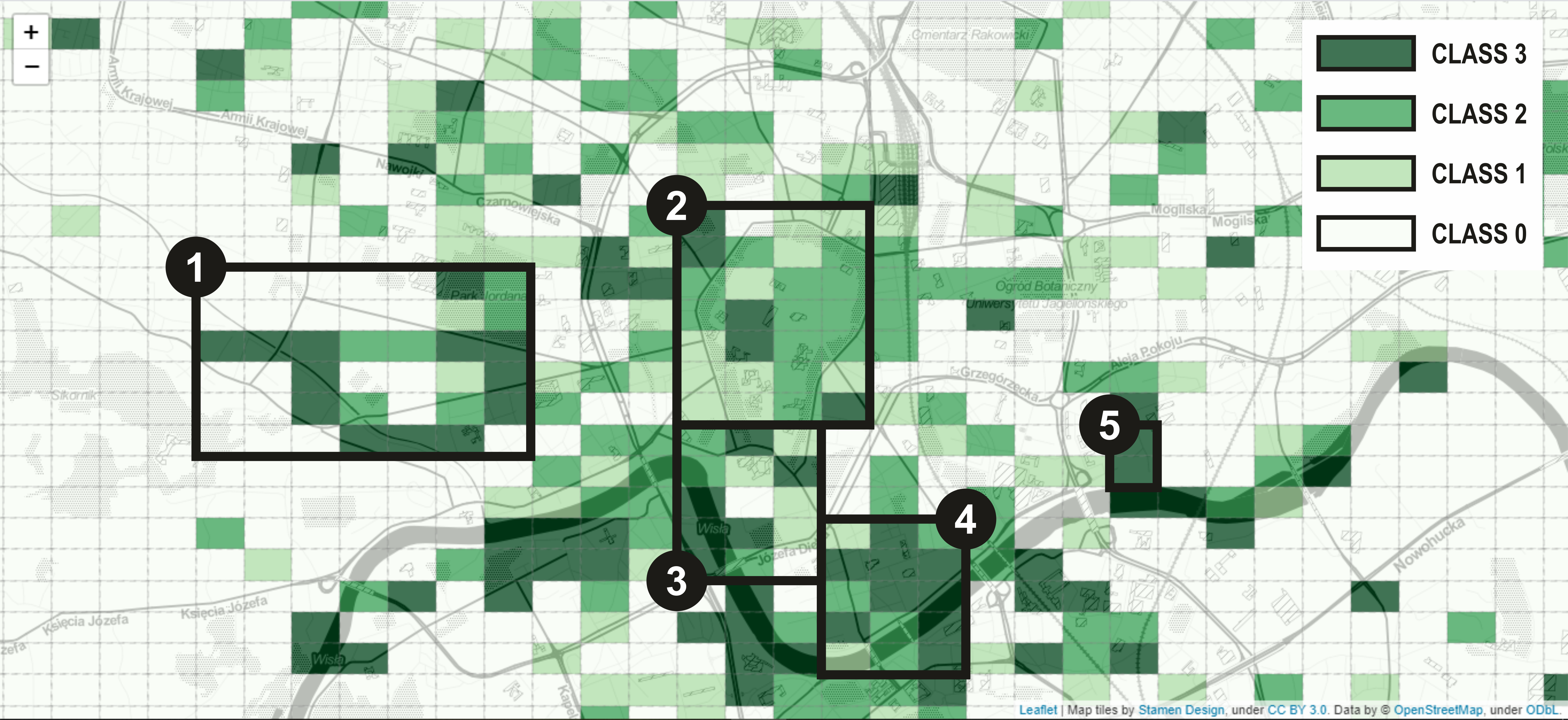}}
\caption{Selected highest-rated spots in Central Kraków: 1. Błonia with the Rudawa Valley and the Jordan Park, 2. the Old Town with the Main Market Square, 3. Wawel Hill with the Vistula Boulevards, 4. Kazimierz with the Vistula Boulevards, 5. the city beach area, highly popular among locals is the new riverside and not listed in official guides. The old town (2) is now more detailed, making the most attractive spots visible.} \label{fig:f7}
\end{figure}

\begin{figure}
\centering
\resizebox*{10cm}{!}{\includegraphics{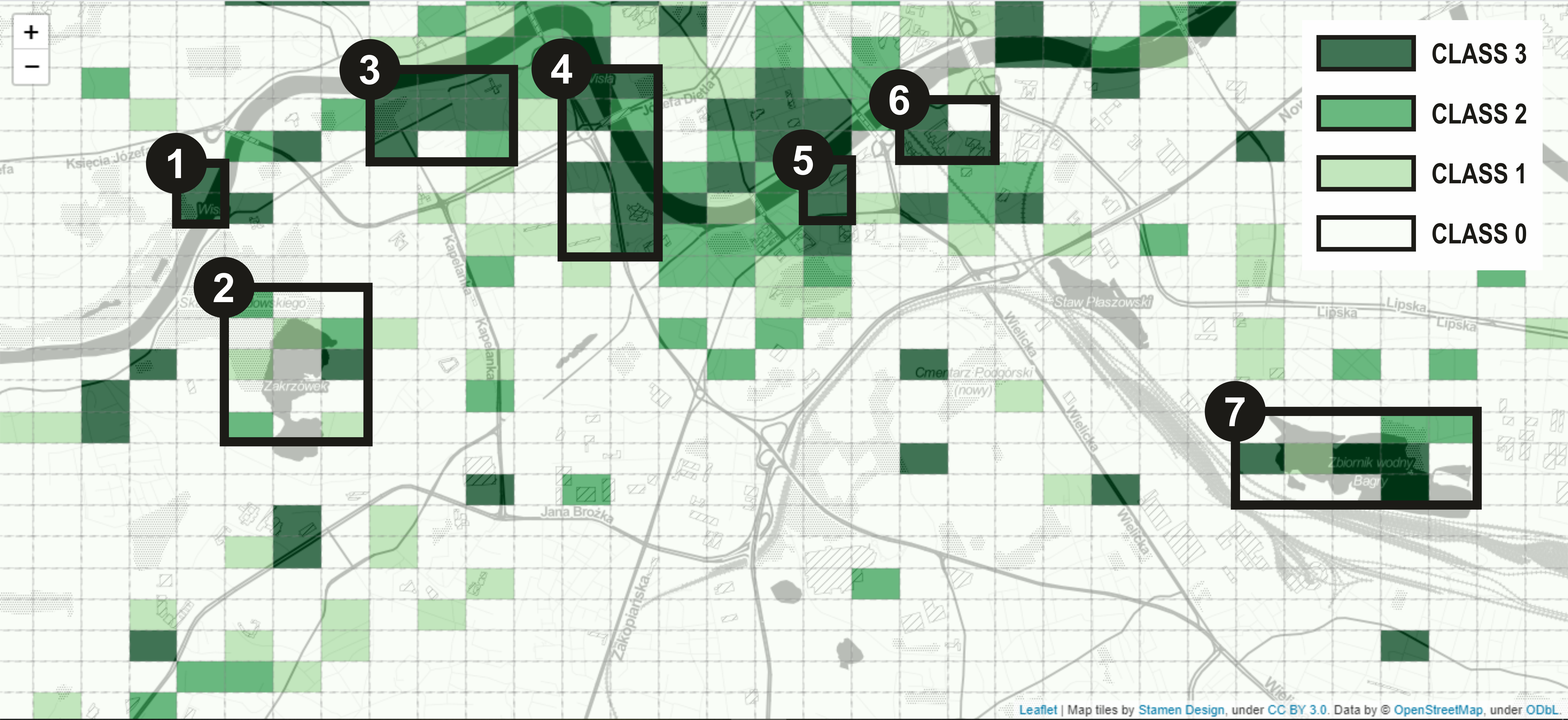}}
\caption{Selected highest-rated spots in South Kraków: 1. the kayak rental station, 2. Zakrzówek lake, 3. the Vistula Boulevards in Stare Dębniki, including the Dębnicki Park, 4. the Vistula Boulevards in Ludwinów, 5. the area of Cricoteka and the Podgórski Market Square, 6. the area of Schindler’s Factory and the MOCAK, 7. Bagry Lake. Pilgrims centre Łagiewniki (to the south), highly ranked in official guides, not identified in our method due to low bike accessibility. The shopping mall at Norymberska street (down from hotspot 2) was identified as attractive, which is clearly a false-positive case that has to be filtered manually.} \label{fig:f8}
\end{figure}

\begin{figure}
\centering
\resizebox*{10cm}{!}{\includegraphics{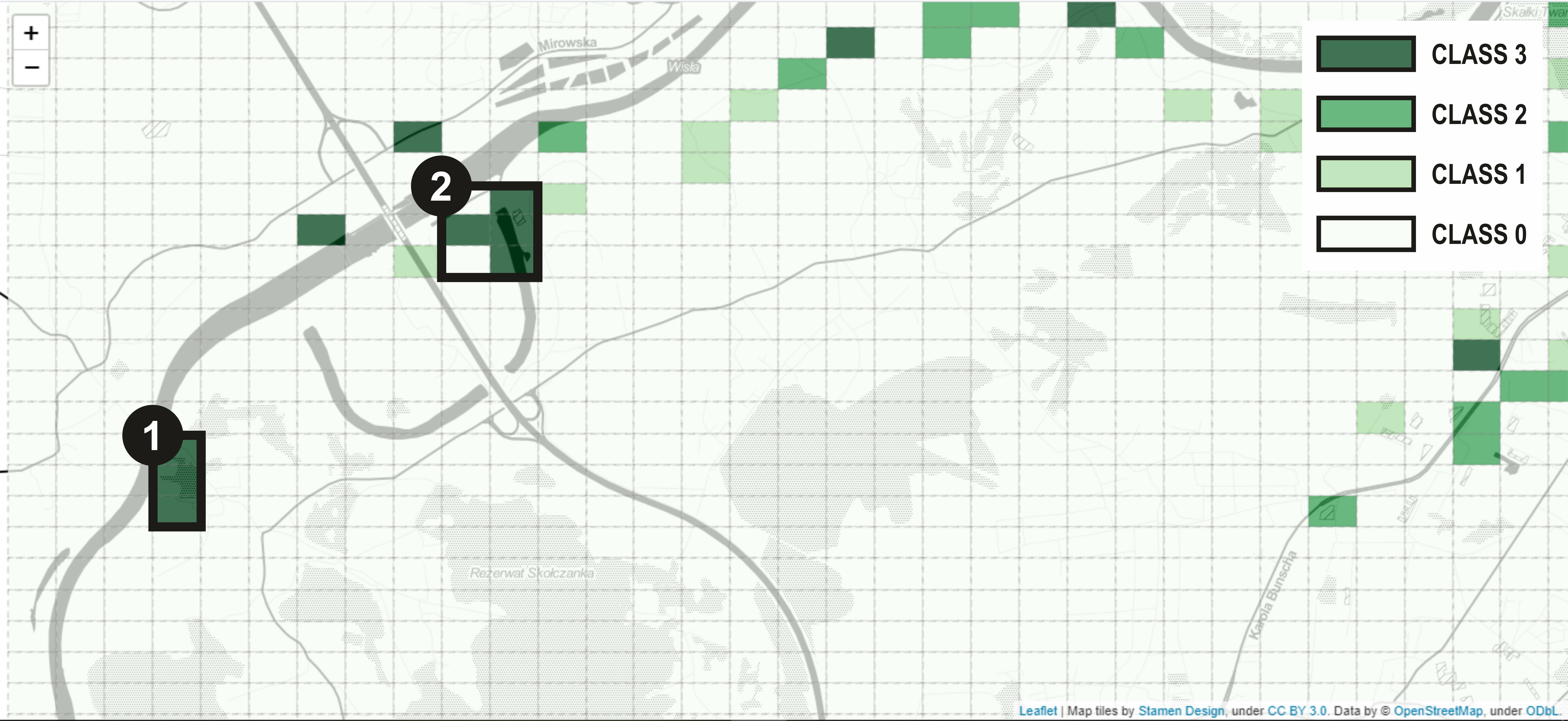}}
\caption{Selected highest-rated spots in West Kraków: 1. the Benedictine Abbey in Tyniec, 2. the kayaking trail in Tyniec. Both attractive yet distant, which leaves a trace of short breaks along the highly popular bike path stretching by the river between old town and Tyniec.} \label{fig:f9}
\end{figure}

\begin{figure}
\centering
\resizebox*{10cm}{!}{\includegraphics{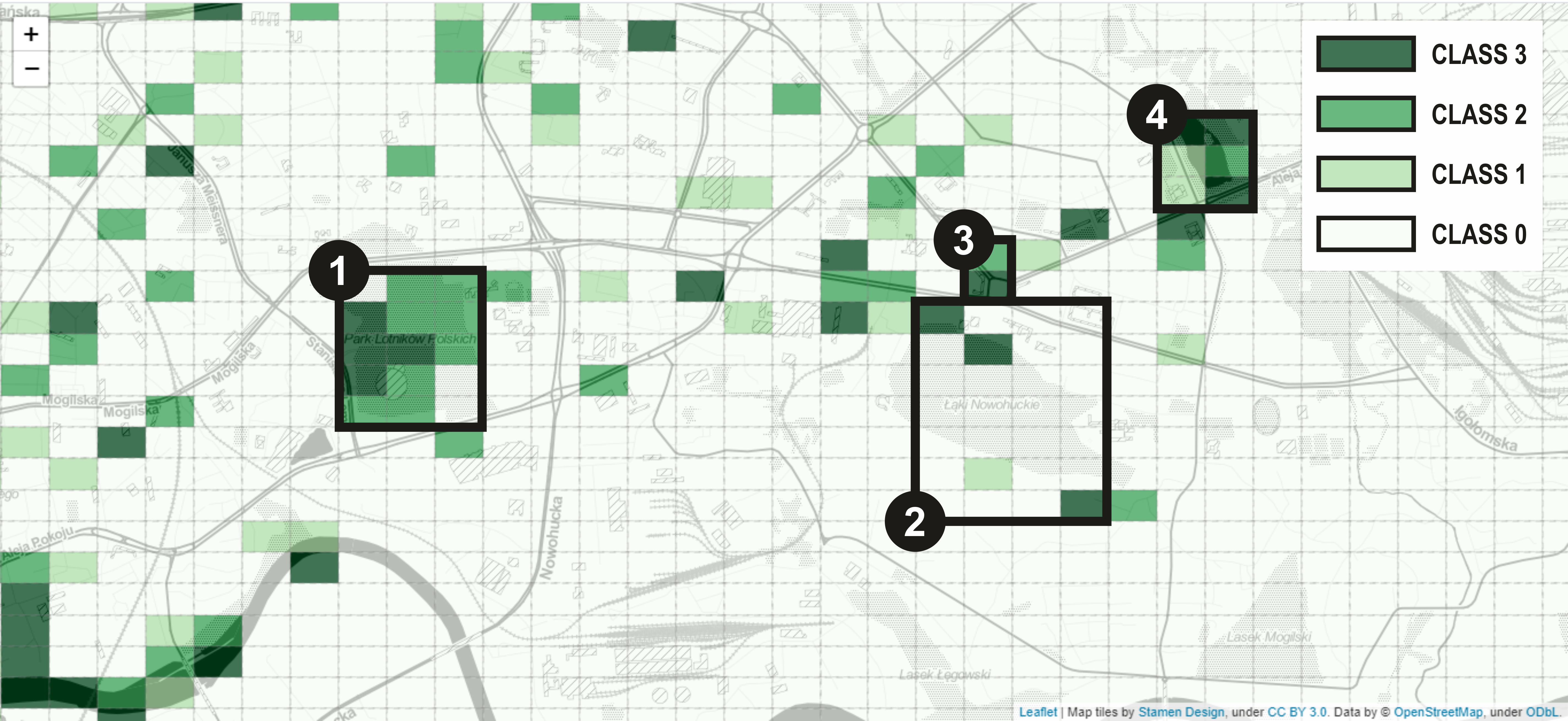}}
\caption{Selected highest-rated spaces in East Kraków: 1. the Polish Aviators’ Park, 2. Nowa Huta Meadows, 3. Centralny Square with Aleja Róż, 4. Nowa Huta Lake. The old Nowa Huta area marked as one equally attractive spot in official guidelines (fig. \ref{fig:f6} ) is now depicted with more detail, making it evident that attractive places concentrate around Plac Centralny and the Lake.} \label{fig:f10}
\end{figure}

Most of the hotspots identified by our method overlap with those mentioned in the literature. Figures \ref{fig:f7} to \ref{fig:f10} zoom in the attractiveness grid in selected areas of Kraków and discuss them. By comparing figure \ref{fig:f6} with figures \ref{fig:f7} to \ref{fig:f10} one can see greater details of hotspots locations and complex, yet clear patterns resulting from our method. 

In table \ref{table3} we compare most important official tourist attractiveness (highly ranked in \cite{faracik2008a}) with our method results. Our method managed to correctly identify all the tourist attractions from the official sources of attractiveness (compare our findings on fig \ref{fig:f5} with official attractiveness on fig \ref{fig:f6}), with two exceptions, both poorly accessible by bike. 

First one is pilgrims centre in Łagiewniki (south), which is typically visited by elder tourists who rarely use city bikes and is poorly accessed by bicycle (as it is located on the hill). The second was the Wolski Forest, with the Zoo and the Piłsudski Mound. While considered as one of the most attractive spots in the city, for topographical reasons, this place is popular among mountain bikers rather than Wavelo users. 

On the other hand, some places classified as attractive with our method, were clearly not touristic shopppiung malls. Shopping can be perceived attractive by tourists and locals, several of shopping malls in Kraków are the attractive ones. They are located near the Old Town (Galeria Krakowska) and Kazimierz (Galeria Kazimierz). However the two examples that we use: the Krokus mall, and shopping centre in Norymberska Street are clearly not attractive and used by locals to supply their basic needs. 
Our method failed to filter them out, yet manual post-processing with a basic background field-knowledge allowed to effectively interpret such false-positive cases.

\begin{table}[]
\centering
\resizebox{\textwidth}{!}{
\begin{tabular}{llllll}

\toprule
\multicolumn{2}{l}{\textbf{Place}}                                                     &  rank: &   &  &                      \\
\multicolumn{2}{l}{}                                                     & official  & our & accuracy${}^*$  & comment                    \\\midrule

                                       & \multicolumn{5}{l}{\textbf{Central} (fig. \ref{fig:f7})}                                                                                                    \\ \midrule
\multicolumn{2}{l}{the Main Market Square}                                    & 3                                    & 3          & TP             &                             \\
\multicolumn{2}{l}{Wawel Castle, and the Vistula   Boulevards} & 3                                    & 3          & TP             &                             \\
\multicolumn{2}{l}{Kazimierz quarter}                                         & 3                                    & 3          & TP             &                             \\
\multicolumn{2}{l}{the city beach in the Courland Boulevard}                  & -                                    & 3          & TP             & recently opened             \\
\multicolumn{2}{l}{Błonia and the Rudawa Valley}                              & 2                                    & 3          & TP             &                             \\
\multicolumn{2}{l}{the Jordan’s Park}                                         & 2                                    & 3          & TP             &                             \\\midrule
                                       & \multicolumn{5}{l}{\textbf{South-west} (fig. \ref{fig:f8}, \ref{fig:f9})}                                                                                              \\ \midrule
\multicolumn{2}{l}{Cricoteka}                             & 1                                    & 3          & TP             &                             \\
\multicolumn{2}{l}{Schindler’s Factory and   the MOCAK}   & 2                                    & 3          & TP             &                             \\
\multicolumn{2}{l}{the Krakus Mound}                                          & 2                                    & 3          & TP             &                             \\
\multicolumn{2}{l}{Zoo and Piłsudski Mound}                                   & 2                                    & 0          & \color{red}{TN}             & poorly accessible by bike   \\
\multicolumn{2}{l}{Bagry Lake}                                                & -                                    & 3          & TP+       & primarily for locals        \\
\multicolumn{2}{l}{the Vistula Boulevards in Stare Dębniki,}                  & 2                                    & 3          & TP             &                             \\
\multicolumn{2}{l}{Pilgrim center Łagiewniki}                                 & 3                                    & 0          & \color{red}{TN}             & poorly accessible by bike   \\
\multicolumn{2}{l}{the Vistula Boulevards in Ludwinów}                        & 2                                    & 3          & TP             &                             \\
\multicolumn{2}{l}{Zakrzówek}                                                 & 1                                    & 3          & TP+      & primarily for locals        \\
\multicolumn{2}{l}{the kayaking trail in Tyniec}                              & 2                                    & 3          & TP             &                             \\
\multicolumn{2}{l}{the Benedictine Abbey in Tyniec}                           & 3                                    & 3          & TP             & distant yet bike-accessible \\
\multicolumn{2}{l}{\textit{Shopping mall at   Norymberska street}}                     & -                                    & 3          & \color{red}{FP}             &   hardly attractive                           \\\midrule
                                       & \multicolumn{5}{l}{\textbf{East} (fig. \ref{fig:f10})}                                                                                                      \\ \midrule
\multicolumn{2}{l}{Centralny Square}                                          & 2                                    & 3          & TP             &                             \\
\multicolumn{2}{l}{the Polish Aviators’ Park}                                 & 1                                    & 3          & TP+       &                             \\
\multicolumn{2}{l}{Nowa Huta Lake}                                            & 2                                    & 3          & TP             &                             \\
\multicolumn{2}{l}{Nowa Huta Meadows}                                         & -                                    & 3          & TP+       & picnic spot for bike trips  \\
\multicolumn{2}{l}{\textit{Krokus   shopping mall}}                                    & -                                    & 3          & \color{red}{FP}            &       hardly not attractive          \\  \midrule   
\end{tabular}}
\caption{Most attractive tourist places according to official listing \cite{faracik2008a} compared with our classifications. Italics denote places either not identified or wrongly classified as attractive our method.
${}^*$TP - true positive, TP+ - added-value to the official sources, TN - true negative, FP - false positive, compare with fig. \ref{fig:matrix2}} \label{table3}
\end{table}


\section{Conclusions and discussion}

We proposed a generic method applicable for any spatiotemporal data from city-bikes, which, since city bikes are nowadays present in most of metropolises worldwide, makes it applicable to explore spatial patterns of stopovers in broad range of cities. The light and replicable method, relying on standard spatiotemporal trip tracks allows to identify hotspots - places with frequent and long stopovers. Which, as we demonstrate are a good proxy of the space attractiveness. By assuming that most of stopovers identified with the method are related to tourism and/or leisure, we identified complex and meaningful spatial patterns, clearly pointing towards city’s most attractive urban hotspots. The results shown that the most frequent stopover locations of Wavelo bike users were, in fact, concentrated in the proximity of the most attractive cultural and natural assets. With the proposed automated manual filtering, not relying on local field knowledge, one can reveal number of valuable findings, both in terms of identifying unknown places as well as quantifying well-known ones. 
While the local knowledge may refine the results and make it reliable indicator of urban attractiveness.

Results of our case-study proved that the method effectively identified most of established Kraków tourist attractions. The recent dynamism in behaviour of tourists, shifting from well-known paths to exploring newly emerging sights, was evident from the emerging patterns. Our method managed to cover it and quantify those changes.

Despite relying on personal and potentially sensitive data, we find the method transparent and ethical. Even though the precise spatial path is recorded, it always starts and ends at the city bike station, rather than at personally sensitive home or workplace. While the users’ ID is not stored in the dataset, his sociodemographic attributes may be disclosed which would enable more detailed analysis differentiating locals from visitors, young from older users, etc. The proposed method has potential in real-time monitoring and can be potentially automated to report the attractiveness and its relative changes over time. Making it an efficient tool for policy makers to monitor shifts in tourist behaviour.

Importantly, in the post-COVID context our method offers an efficient and inexpensive monitoring framework allowing to understand how the pandemic changes impacted the consumption of urban space by tourists, locals and visitors. Allowing to quickly identify most visited, possibly crowded, places where intervention may be needed to stop virus-spreading.

Obviously, our method has some limitations. It relies on sequence of filters, where thresholds need to be manually parameterized (e.g., cut-off distance from traffic lights, or from station). While it effectively filters commuting trips, as long as user type remains unknown, the leisure remains indistinguishable from tourism, and local residents from visitors. This shall be further enhanced with labelled data. While the city bike systems are often limited in space and rental stations are not evenly distributed everywhere in the city. This was not the case for Kraków, yet in cities where coverage is not complete, this may obscure the overall image and fail to yield a complete spatial pattern. 

Finally, to verify the results before drawing conclusions a basic field knowledge is needed. In the case of Kraków, some definitely non-attractive shopping malls were misclassified with our method. Yet in any case, virtual or physical site visit may always verify its actual attractiveness. For instance, for us, the locals of Kraków, the data revealed the place that we were not aware of (hidden garden at Karmelicka street).

We believe that with this method we fill some of the gaps in research on the urban space attractiveness for residents and tourists.

\paragraph*{Acknowledgement(s)}

We thank the City of Kraków, the operator of Wavelo city bike for the data for analysis.









\bibliographystyle{apacite}
\bibliography{main.bib}

\end{document}